\documentclass[aps,prd,reprint,showpacs,showkeys,amsmath,amssymb,amsfonts]{revtex4-2}
\pdfoutput=1
\usepackage{blindtext}
\usepackage{graphicx}
\usepackage{fullpage}
\usepackage{dcolumn}
\usepackage{tabularx}
\usepackage{natbib}
\usepackage{mathtools}

\usepackage[x11names]{xcolor}
\usepackage{hyperref}
\hypersetup{linkcolor=blue, citecolor=cyan, filecolor=cyan, urlcolor=blue, linktoc=all}

\begin{document}
\title{Spin Tetrad Formalism Of Circular Polarization States In Relativistic Jets}
\author{Ronald S. Gamble, Jr}
\affiliation{NASA Goddard Space Flight Center\\ Greenbelt, MD\\}
\affiliation{Center for Research and Exploration in Space Science and Technology}
\affiliation{Department of Astronomy, University of Maryland\\College Park, MD}
\date{2022}
\email{ronald.s.gamble@nasa.gov}

\begin{abstract}
Relativistic jets from active galactic nuclei (AGN) have been of peak interest in the high-energy astrophysics community for their uniquely dynamic nature and incredible radiative power; emanating from supermassive black holes and similarly accreting compact dense objects. An overall consensus on relativistic jet formation states that accelerated outflow at high Lorentz factors are generated by a complex relationship between the accretion disk of the system and the frame-dragging effects of the rotating massive central object. This paper will provide a basis for which circular polarization states, defined using a spin tetrad formalism, contribute to a description for the angular momentum flux in the jet emanating from the central engine. A representation of the Kerr spacetime with positive cosmological constant background is used in formulating the spin tetrad forms. A discussion on unresolved problems in jet formation and how we can use multi-method observations with polarimetry of AGN to direct future theoretical descriptions will also be given. 

\end{abstract}

\maketitle


\section{Introduction} \label{sec: intro}
\subsection{Review}
AGN are quite possibly the largest, most luminous, extragalactic objects observed in the known universe. The FERMI-LAT collaboration has generated an extensive catalogue of AGN in the high-energy regime\cite{fermi10yr, Abdollahi2020FermiLA, fermi2020, TeV1ES_2020, 2016ApJS.226...20F}. A number of probes and missions are dedicated to the multi-messenger aspects of observing these energetic objects with variable emission \cite{amego, ixpe}. The All-sky Medium Energy Gamma-ray Observatory (AMEGO) \cite{amego} is focused on the medium energy scaled x-ray sky and includes a variety of methods to detect and observe polarimetry from MeV blazars at very high redshift. Currently, the mechanism for relativistic jet emission associated with active galactic nuclei, and other high-energy astrophysical objects like gamma-ray bursts, and microquasars, has been of current interest in the astrophysics scientific community. Jet formation theory and emission is a major problem yet to be solved in high-energy astrophysics, with extensive reviews given describing the properties of jets \cite{Harris2006, Blandford2019} . One of the most widely argued models for describing this type of emission has been the Blandford-Znajek (BZ) process\cite{bz77}. This process describes the rotational energy extraction from black holes involving the torsion of magnetic field lines resulting in Poynting flux dominated outflows parallel to the rotation axis of the central object \cite{bz77} and \cite{znajek77}. Developed theories and models of matter-energy ejection from a central engine feature the luminosity function below as a template for comparing magnetic field contributions to the power emitted,
\begin{equation}\label{bzlumn}
L_{BZ} = f(\alpha_{H}) B_{\phi}^{2} r^{2}_{s} (c 8\pi)^{-1}
\end{equation}
The nature of such highly complex energetic emission mechanisms from these systems, which feature event horizons in rotating spacetimes, has been studied extensively over the last few decades \cite{YKim:2020cv, Kawashima2020, Doeleman:2012vd, Williams1995, Williams2004, king_2021}. Recent numerical models incorporating MHD and GRMHD methods have shown that a major contribution to jet outflows are from the poloidal magnetic field configurations from relativistic matter accreting on to the central object \cite{koide2020, Contopoulos2014, Komissarov_2005}. 
\begin{figure}[h]
\centering
\includegraphics[scale=0.2]{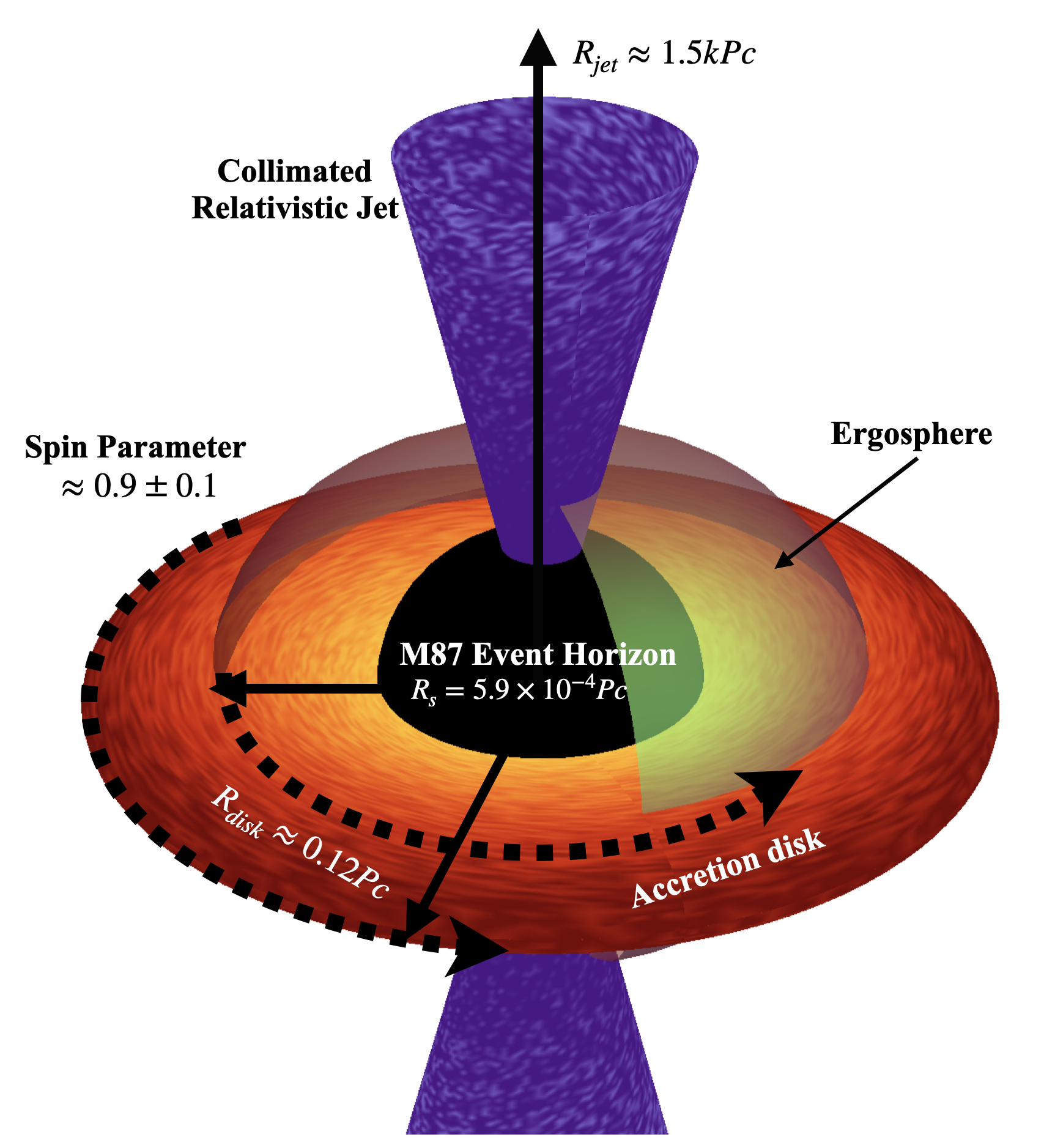}
\caption{Visualization of disk-jet model of M87 and the associated relativistic jet to scale. Where the Schwarzschild length scale is $R_{s}=5.9\times 10^{-4 } Pc$, disk radius of $R_{disk} \approx 0.12 Pc$ and jet length of $R_{jet} \approx 1.5 kPc$.} 
\end{figure}
Unanswered questions on the relativistic nature of these jets involve figuring out how particles that make up the jet's content are accelerated to ultra-relativistic speeds whose Lorentz factors are $\Gamma_{Lorentz} >5$. What is the origin of the relativistic particles that produce non-thermal radiation we observe? And how do these jets become \textit{matter loaded}? Focusing on the theoretical aspects of jet formation mechanisms and high-energy polarizations in extreme gravity environments \cite{Rani2019}, there are still fundamental questions that continue to remain unresolved.
Recent observations of optical circular polarizations in blazar jets\cite{cp_blazar2021, Rieger2005} present a new regime for probing the physics and material content in high-energy astrophysical environments. Another, less explored problem with jets, is the causal connection of the jet to the exterior Kerr spacetime. A discussion on the causal connection of BZ emission as a flavor of a Penrose-type emission can be seen in \cite{komissarov2004}. An application of the emission processes to alternative or extensions of relativity to include other field-theoretic descriptions of spacetime\cite{altgr_bz} and the incorporation of spinor representation in \cite{stokesspinor} has shown the versatility in the decades-old theory but, again, exhibits how the BZ-process needs extensions to incorporate the sources of the magnetic fields it describes as argued in\cite{king_2021, bzcurrent2}.

The remaining flow of this paper is as follows. Section \ref{GR} provides an overview of the importance of incorporating a broader contribution from general relativity in multi-messenger characteristics of AGN, focusing in on the non-thermal properties characterizing AGN and specifically using blazars as a primary type of source. The succeeding section \ref{kerr} provides a brief description of the geometry of a curved spacetime encompassing an AGN, as given by a kerr spacetime metric. This section is then furthered by an introduction to a \textit{spin tetrad} formalism for null propagating fields, ending with a geometric interpretation of the spin angular momentum coupling between gravitational and electromagnetic fields surrounding a supermassive black hole (SMBH) as a compact host.
\\

\subsection{Incorporating General Relativistic Theory In Jet Physics}\label{GR}

A lot can be revealed regarding the phenomena occurring in the environments of AGN using general relativity as a baseline theory. Utilizing this as segway into the subsequent sections, we explore why relativity theory is needed to understand the physical mechanisms taking place in the environment and evolution of AGN. There are numerous instances where formulations of jet emission require geometric constructs from General Relativity, especially when near-horizon jet launching environments are considered. More importantly, the work presented in this manuscript is focused on high-energy jets from supermassive black holes (AGN). Almost all matter-energy emission mechanisms intrinsically require the use of general relativity \cite{bz77, znajek77, altgr_bz, Williams1995, Williams2004, king_2021, bzcurrent2}, most importantly, relativistic jets from AGN are those that come from immense sources of gravity due to the supermassive black holes that generate them. With this statement in mind, we now move to describe the radiative characteristics of relativistic jets in the setting of a rotating black hole geometry (the Kerr spacetime). This will allow one to describe the radiation observed from jets within the constructs of a 4-dimensional spacetime algebra, incorporating the corresponding mechanics of black holes within the theory.
\begin{figure}[h]
\centering
\includegraphics[scale=0.25]{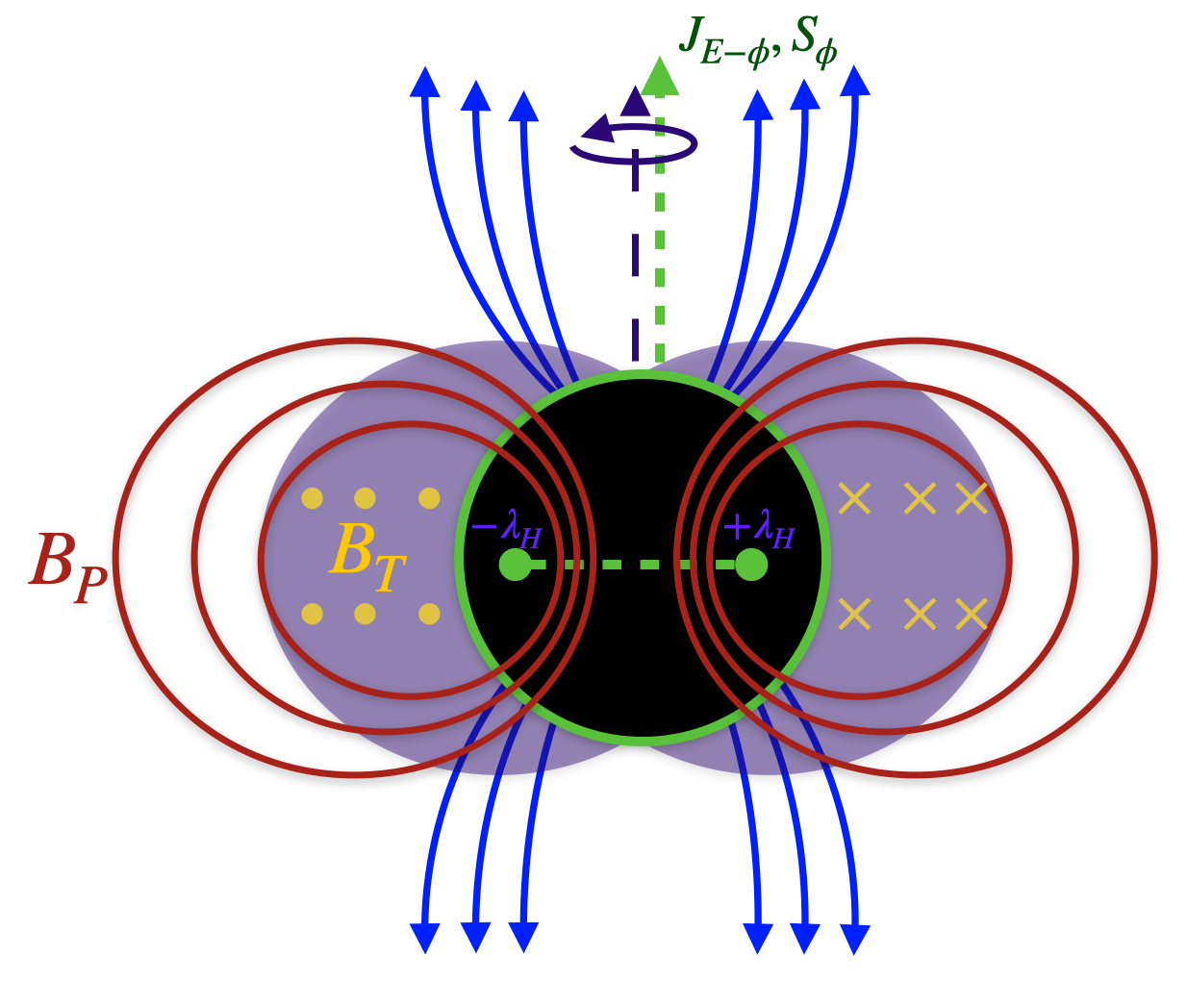}
\caption{Geometric representation of the poloidal $B_{p}$, and toroidal $B_{T}$, magnetic fields superimposed on the horizon + ergosphere structures. The current associated with this configuration is denoted as $J_{E-\phi}$, parallel with the black hole spin axis and jet poynting vector $S_{\phi}$ in the $\hat\phi$ direction. The horizon foci located at $\lambda_{H} = \pm \frac{1}{\sqrt{3}}r_{s}$ \cite{rgdiss} identify a proposed inner boundary to magnetic field lines threading the horizon surface. }
\end{figure}
It is widely understood that the nature of such energetic processes require a sufficiently strong relationship between the magnetic field structure and the torsion of spacetime, represented as frame-dragging in the local frame of the black hole. the figure above illustrates this geometric description of the black hole-Jet system.

General relativity can be a powerful tool for deciphering the dynamic motion of matter and the physics that influences the observed motion. In light of current efforts in observing the degree of polarization in the relativistic jets of high-energy AGN, incorporating geometric descriptions of curved spacetime diversifies formulations of jet emissions and particle phenomenology. 
\\
\section{Geometric Representation of AGN}\label{kerr}
\subsection{Kerr Geometry}
As an accurate description of the spacetime surrounding a central compact massive object in the high-energy density regime, we utilize the Kerr metric of spacetime surrounding such an object\cite{kerr}. Considering the Kerr metric as the spacetime describing the central SMBH in this system.
\begin{widetext}
\begin{eqnarray}
ds^2_{Kerr}&=&\left(\frac{r_s r}{\rho^2}\right)c^{2}dt^{2}-\left(\frac{\rho^2}{\Delta}\right)dr^{2}-\rho^2 d\theta ^2\\
&-&\left(r^2+\alpha ^2+\frac{r_s r\alpha^2}{\rho^2}\sin^2\theta \right)\sin^2\theta d\varphi^2+\frac{2r_s r\alpha sin^{2}\theta}{\rho^2}cdtd\varphi \nonumber
\end{eqnarray}
\end{widetext}
For brevity we define the functions, $\Delta, \rho$ as $\rho = r^{2}+\alpha^{2}Cos^{2}\theta$ and
$\Delta = r^{2} -r_{s}r +\alpha^{2}$, respectively. In line with this, the black hole spin parameter is defined as $\alpha = J/M_{H}c$, with the angular velocity of the horizon as
\begin{equation}
\omega = -\frac{g_{t\phi}}{g_{\phi\phi}} = \frac{r_{s}\alpha r c}{\rho(r^{2} + \alpha^{2})+ r_{s}r\alpha^{2}\sin^{2}\eta}
\end{equation}

\subsection{Spin Tetrad Formalism}\label{spintetrad}
A systemic way of projecting geometric objects into spin components is to work in a \textit{spin tetrad}. The spin tetrad representation describes how an object transforms under a projection onto a preferred spin axis. In the case of electromagnetic and gravitational waves, the preferred axis is parallel with an associated Poynting vector for the waves. 

Beginning with a Boyer-Lindquist tetrad $\{\gamma_{t},\gamma_{r},\gamma_{\theta},\gamma_{\phi}\}$ in the Kerr spacetime with the preferred spin axis in the $\hat\phi$-direction, the spin tetrad axes are then defined as complex combinations of the transverse axes $\{\gamma_{+},\gamma_{-}\}$,
\begin{subequations}
\begin{align}
\gamma_{+} &\equiv \frac{1}{\sqrt{2}}\left(\gamma_{r} + i\gamma_{\theta}\right)\\
\gamma_{-} &\equiv \frac{1}{\sqrt{2}}\left(\gamma_{r} - i\gamma_{\theta}\right)
\end{align}
\end{subequations}
Where the metric of the spin tetrad $\{\gamma_{t}, \gamma_{\phi},\gamma_{+},\gamma_{-}\}$ can be written as
\begin{equation}
\gamma_{LK} =\left(
\begin{array}{cccc}
 -1 & 0 & 0 & 0 \\
 0 & 1 & 0 & 0 \\
 0 & 0 & 0 & 1 \\
 0 & 0 & 1 & 0 \\
\end{array}
\right).
\end{equation}
Given this definition of the spin tetrad, the spin axes $\{\gamma_{+},\gamma_{-}\}$ are intrinsically null $\gamma_{+} \cdot \gamma_{+} = \gamma_{-} \cdot \gamma_{-} = 0$, where as their inner product with respect to each other are nonzero $\gamma_{+} \cdot \gamma_{-} = 1 $. Under a right-handed rotation by an angle $\psi$ about the preferred axis $\gamma_{\phi}$, the transverse axes $\gamma_{r}, \gamma_{\theta}$ transform as
\begin{subequations}
\begin{eqnarray}
\gamma_{r} &\to cos(\psi)\gamma_{r} + sin(\psi)\gamma_{\theta}\\
\gamma_{\theta} &\to sin(\psi)\gamma_{r} - cos(\psi)\gamma_{\theta}
\end{eqnarray}
\end{subequations}
It follows that the spin axes $\{\gamma_{+} \text{and}\gamma_{-}\}$ transform under a right-handed rotation about the orthogonal axis $\gamma_{\phi}$ as
\begin{equation}
\gamma_{\pm} \to e^{\mp i\psi s}\gamma_{\pm}
\end{equation}
This transformation identifies the spin axes $\gamma_{+},\gamma_{-}$ as having (+) or (-) helicity respectively. An object can be identified as having \textit{spin} if it varies by $e^{\mp\mathit{i}\psi s}$. An interesting concept arises due to this definition of spin, in a mathematical and physical sense, that spin may be a more global property attributed to the angular momentum flux from a rotating black hole featuring a relativistic jet.

With this in mind, we move to use this representation for describing the local spacetime frame the is dragged exterior to the event horizon at $z_{H} =\frac{1}{2}\left[r_{s} - \sqrt{r^{2}_{s} -4\alpha^{2}}\right]$. Using the stated B-L tetrad, we can write the frame metric in terms of the orthogonal set of veirbien 1-forms, $e^{m}_{\nu}dx^{\nu}$
\begin{subequations}
\begin{eqnarray}
\gamma^{0} &=& e^{0}_{\nu}dx^{\nu} = \left[\left(\frac{r_s r}{\rho^2}\right)c^{2}\right]^{1/2}dt\\
\gamma^{1} &=& e^{1}_{\nu}dx^{\nu} = -\left[\frac{\rho^2}{\Delta}\right]^{1/2} dr\\
\gamma^{2} &=& e^{2}_{\nu}dx^{\nu} = \left[-\rho^2\right]^{1/2} d\theta\\
\gamma^{3} &=& e^{3}_{\nu}dx^{\nu}\\ 
&=& \left[-\left(r^2+\alpha ^2+\frac{r_s r\alpha^2}{\rho^2}\sin^2\theta \right)\sin^2\theta\right]d\varphi \nonumber
\end{eqnarray}
\end{subequations}
\\


%
%
%

\section{Spin Polarization in the Kerr Spacetime}
\subsection{Polarization Tensor of a Null Field}
To accurately construct a description of the polarization of emitted radiation from high-energy sources, we must derive a description of the Stokes Parameters in the setting of general relativity such that these parameters can then be used in the subsequent formulation of the spin and orbital angular momentum carried by these point sources described in previous sections. With this in mind we follow a brief derivation of the stokes parameters, carefully presented in \cite{grav_sp1}, as applied to a spin tetrad frame with a kerr background.

A solution to the electromagnetic field equations can be given as the real part of the Maxwell-Faraday tensor as
\begin{equation}
F_{\alpha\beta} = \mathrm{Re}\{U_{\alpha\beta}e^{\mathit{i}\phi}\}
\end{equation}
Where the field tensor is composed of a function of position in $U_{\alpha\beta}(x)$ and phase $\phi(\theta)$ to which we can define the null vector $k_{\alpha}\equiv \partial_{\alpha}\phi$ with properties: (i) $k_{\alpha}k^{\alpha}=0$, (ii) $\mathcal{D}k_{\alpha}=0$. Where $\mathcal{D}$ is the directional derivative
\begin{equation}
\mathcal{D}\equiv \frac{D}{D\lambda} = \frac{dx^{\mu}}{d\lambda}\nabla_{\mu}
\end{equation}
with $\lambda$ as an affine parameter for parallel propagation of the field. $F_{\alpha\beta}$ is then considered a null field satisfying the equation 
\begin{equation}
F_{\alpha\beta}k^{\alpha}=0
\end{equation}
with a corresponding propagation law given as
\begin{equation}
\frac{dx^{\alpha}}{d\lambda}\nabla^{\alpha}F_{\alpha\beta} - 2\left[-\frac{1}{2}\nabla_{\alpha}k^{\alpha}F_{\alpha\beta}\right] = 0
\end{equation}
in the Lorentz gauge. We can then proceed to describe this representation of the electromagnetic field tensor in terms of the NP-formalism using a null tetrad. The electromagnetic field tensor then has a complete representation as
\begin{equation}
F_{\alpha\beta} = k_{[\alpha}m_{\beta]} + \mathit{i}k_{[\alpha}\overline{m}_{\beta]}
\end{equation}
With this we can write down an electromagnetic polarization tensor, $J_{\alpha\beta\gamma\sigma}$
\begin{equation}
J_{\alpha\beta\gamma\sigma} = \frac{1}{2}<F_{\alpha\beta}\overline{F}_{\gamma\sigma}>
\end{equation}
where the $<\dots>$ represent the average over many periods of the propagating wave. The polarization tensor then has similar antisymmetric and conjugate properties to that of the field tensor, 
\begin{subequations}
\begin{eqnarray}
J_{\alpha\beta\gamma\sigma} = J_{[\alpha\beta][\gamma\sigma]} = \overline{J}_{\gamma\sigma\alpha\beta}\\
g^{\alpha\beta}g^{\gamma\sigma}J_{\alpha\beta\gamma\sigma} = J_{\alpha\beta\gamma\sigma}k^{\beta}=0
\end{eqnarray}
\end{subequations}
In the rest frame of an observer (i.e. detector, probe, etc.) $\mathcal{P}$ with 4-velocity, $u^{\alpha}$, the polarization properties of radiation are expressed using the polarization tensor
\begin{equation}
J_{\alpha\beta} = J_{\alpha\beta\gamma\sigma}u^{\gamma}u^{\sigma}
\end{equation}
Since the components of the electric field are given by $E^{\alpha} = F^{\alpha\beta}u_{\beta}$, for $u_{\beta} = \left(u_{t},0,0,0\right)$.
Where, 
\begin{equation}
J_{\alpha\beta} = <E_{\alpha}\overline{E}_{\beta}>
\end{equation}
satisfies the property $J_{\alpha\beta}u^{\beta}= J_{\alpha\beta}k^{\beta}=0$. Thus, the average energy-momentum of the field $F_{\alpha\beta} = \mathrm{Re}\{U_{\alpha\beta}e^{\mathit{i}\phi}\}$ is
\begin{equation}
<T_{\mu\nu}> = \frac{1}{4}\left(<F_{\mu\gamma}\overline{F}_{\nu}{}^{\gamma}> + <F_{\mu\gamma}\overline{F}_{\gamma}{}^{\nu}>\right).
\end{equation}
With respect to the polarization tensor, the energy is given by the expression
\begin{equation}
<T_{\mu\nu}>u^{\mu}u^{\nu} = J_{\gamma\mu}{}^{\gamma}{}_{\nu}u^{\mu}u^{\nu} = J_{\gamma}{}^{\gamma} = J
\end{equation}
This states that the average energy measured by an observer or detector at $\mathcal{P}$ is given by an invariant of the radiation polarization tensor, agreeing with the traditional formulation in a specified coordinate basis. This radiation polarization tensor represents the null characteristics of a polarized signal with respect to the electric field. This tensor has suitability in further defining the complete state of polarization of the signal in terms of the Stokes parameters (SP) within a general relativistic setting. 
\\
\subsection{Stokes Parameters In The Spin Tetrad}
\begin{figure}[h]\label{poincare}
\centering
\includegraphics[scale=0.25]{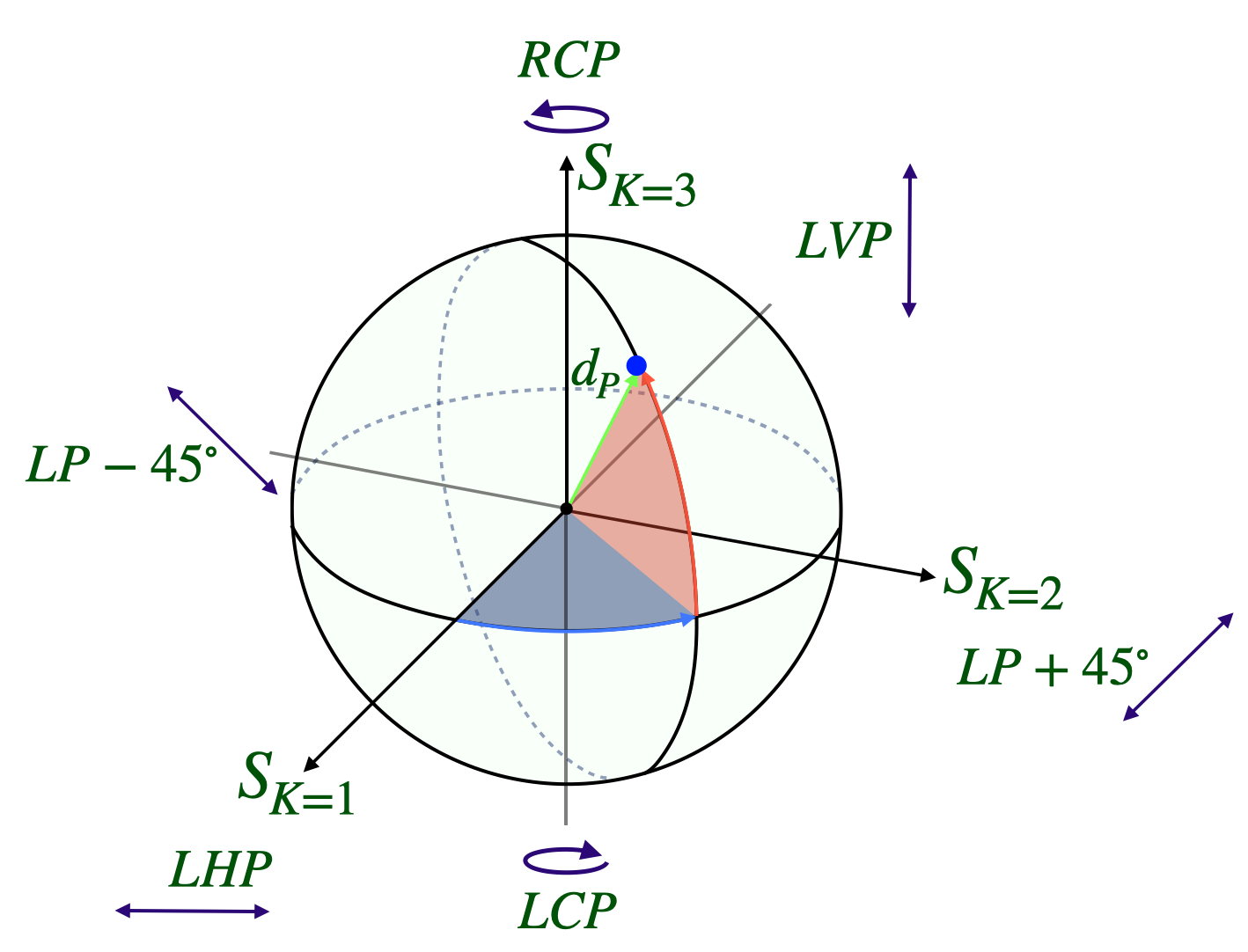}
\caption{Schematic diagram of polarization states projected on to the Poincare sphere. Where the right circular (RCP), left circular (LCP), linear $\pm 45^{\circ}$ (LP$\pm45^{\circ}$), and the linear horizontal (LHP) and vertical (LVP) polarization states are visually represented.} 
\end{figure}

Introducing a spin tetrad in this description for the polarization states of propagating signal provides a rather interesting and intuitive definition of the SP as they relate to a system with active flux of angular momentum; AGN. The null characteristic of the spin tetrad makes it well-suited for describing null fields such as electromagnetism that propagate at the speed light. With this mechanism in place, we can describe the null characteristics of the electromagnetic field in a rotating spacetime with axial symmetry using the spin tetrad stated in section \ref{spintetrad}. Under this formalism it is then intuitive to redefine the $\gamma_{+}$ and $\gamma_{-}$ tetrad components as complex conjugates of each other
\begin{equation}
\gamma_{-} \equiv \overline{\gamma}_{+}
\end{equation}
Here the $\gamma_{-}$ vector is given as the complex conjugate of the $\gamma_{+}$ vector in this formalism. Transforming to this new tetrad frame provides for a better description of the symmetries attributed to the polarization state of the electric field. For brevity the SP in a flat spacetime representation are given as
\begin{subequations}
\begin{eqnarray}
S_{0} &=& I = E^{2}_{x} + E^{2}_{y}\\
S_{1} &=& Q = E^{2}_{x} - E^{2}_{y}\\
S_{2} &=& U = 2E_{x}E_{y}cos(\delta)\\
S_{3} &=& V = 2E_{x}E_{y}sin(\delta)
\end{eqnarray}
\end{subequations}
Where the angle $\delta$ refers to the angle between the left-handed polarization (LHP) $S_{2}$ and the right-handed polarization (RHP) $S_{3}$ projected on to the Poincare sphere. A rotation of any of the SP components by an angle of $180^{\circ}$ leaves them invariant as projected on to the poincare sphere. The radius of this spherical projection is given by the inequality for the most general case of a polychromatic signal $I^{2} \geq Q^{2} + U^{2} + V^{2}$ visualized in figure \ref{poincare}.

The degree of total polarization is then given by
\begin{equation}
d_{P} = \frac{\sqrt{Q^{2} + U^{2} +V^{2}}}{I}
\end{equation}
to which we can also define a degree of circular polarization as
\begin{equation}
d_{C} = \frac{\sqrt{V^{2}}}{I} 
\end{equation}
and degree of linear polarization
\begin{equation}
d_{L} = \frac{\sqrt{Q^{2}+U^{2}}}{I}.
\end{equation}
If we then consider the same observer at $\mathcal{P}$ who sees the polarized signal propagating in the local z-direction using a spin tetrad $e_{A}^{\mu} = \{ \mathit{l}^{\mu}, \mathit{n}^{\mu}, \gamma^{\mu}_{+}, \gamma^{\mu}_{-}\}$, where the capitalized roman letters label the tetrad frame and greek letters label the coordinate frame. We can thus make the substitution for the complex conjugation of the $\gamma^{\mu}_{+}$ vector as $ \gamma^{\mu} =\gamma^{\mu}_{+}$ and $ \overline{\gamma}^{\mu} =\gamma^{\mu}_{-}$. The vectors of this tetrad frame are then defined as the set of axes
\begin{eqnarray}
\mathit{l}^{\mu} &=& \frac{1}{\sqrt{2}}(e_{t} + e_{\phi})^{\mu}\\
\mathit{n}^{\mu} &=& \frac{1}{\sqrt{2}}(e_{t} - e_{\phi})^{\mu}\\
\gamma^{\mu} &=& \frac{1}{\sqrt{2}}(e_{r} + \mathit{i}e_{\theta})^{\mu}\\
\overline{\gamma}^{\mu} &=& \frac{1}{\sqrt{2}}(e_{r} - \mathit{i}e_{\theta})^{\mu}\
\end{eqnarray}
retaining the property of lying along null trajectories for $\mathit{l}^{\mu}\mathit{n}_{\mu} =-\gamma^{\mu}\overline{\gamma}_{\mu} =1$ and following the transformation using the NP spin tetrad metric, $\gamma_{LK}$, where
\begin{equation}
\gamma_{LK} =\frac{1}{\sqrt{2}}\left(
\begin{array}{cccc}
 1 & 0 & 0 & 1 \\
 1 & 0 & 0 & -1 \\
 0 & 1 & \mathit{i} & 0 \\
 0 & 1 & -\mathit{i} & 0 \\
\end{array}
\right).
\end{equation}
Now it is easy to see that we can write down the SP in this frame, capturing information about the geometric properties of spin as it relates to the polarization states. The SP in this frame in terms of the electric field components can be written as
\begin{subequations}
\begin{eqnarray}
S_{0} &=& \left<\left|E_{1}\right|^{2} + \left|E_{2}\right|^{2} \right>\\
S_{1} &=& \left<\left|E_{1}\right|^{2} - \left|E_{2}\right|^{2} \right>\\
S_{2} &=& \left< E_{1}\overline{E}_{2} + \overline{E}_{1}E_{2} \right>\\
S_{3} &=& \mathit{i}\left< E_{1}\overline{E}_{2} - \overline{E}_{1}E_{2} \right>
\end{eqnarray}
\end{subequations}
Thus we can transform this set of SP using the polarization $J_{\alpha\beta} = <E_{\alpha}\overline{E}_{\beta}>$,
\begin{subequations}\label{sptetrad}
\begin{eqnarray}
S_{0} &=& J_{\alpha\beta}\left(e^{\alpha}_{1}e^{\beta}_{1} + e^{\alpha}_{2}e^{\beta}_{2} \right)\\
S_{1} &=& J_{\alpha\beta}\left(e^{\alpha}_{1}e^{\beta}_{1} - e^{\alpha}_{2}e^{\beta}_{2} \right)\\
S_{2} &=& J_{\alpha\beta}\left(e^{\alpha}_{1}e^{\beta}_{2} + e^{\alpha}_{2}e^{\beta}_{1} \right)\\
S_{3} &=& \mathit{i}J_{\alpha\beta}\left(e^{\alpha}_{1}e^{\beta}_{2} - e^{\alpha}_{2}e^{\beta}_{1} \right).
\end{eqnarray}
\end{subequations}
With this set of SP within the framework of a null tetrad we can move towards formulating a formal relationship between the SP and the flux of angular momentum. The SP will prove to be a valuable tool in a description of the spin and orbital angular momentum of polarized states as it relates back to the rotation of the black hole.
\\
\section{Angular Momentum Flux Of Polarized States}\label{sec: amflux}
\subsection{Spin Angular Momentum Of Polarization States}
For a beam centered at the origin $\theta = \pi/2, r = 0$ with respect to the axis of rotation for the jet, parallel to the Poynting vector of propagation, the orbital angular momentum (OAM) will be zero. Any deviation from this symmetry, or precession, should then display a nonzero OAM density dependent on the distance from symmetry deviation in a standard coordinate frame as
\begin{equation}\label{oam}
\frac{1}{\sqrt{g}}\frac{dL_{OAM}}{d^{\mu}r} = \epsilon_{0}\left( cF^{0\mu}\left(r_{[\alpha}\nabla_{\alpha]}\right) A_{\mu}\right)
\end{equation}
This expression states that the electromagnetic field of the beam will carry a nonzero OAM that originates external to the optical field of the beam. Thus, providing insight to the external physics taking place in the black hole environment orthogonal to the rotation of the event horizon at $r = z_{H} = \frac{1}{2}\left[r_{s} - \sqrt{r^{2}_{s} -4\alpha^{2}}\right]$. The same can be said for the Spin Angular Momentum (SAM) in 
\begin{equation}\label{sam}
\frac{1}{\sqrt{g}}\frac{dL_{SAM}}{d^{\mu}r} = \epsilon_{0}cF_{[0\mu}A_{\mu]}
\end{equation}
The component of angular momentum of light associated with the quantum spin and rotation between the polarization degrees of freedom of the photon as an elementary particle.
And thus a \textit{total angular momentum density} $\mathcal{J} = L_{SAM} + L_{OAM}$ can be defined as the sum of the two.
\begin{eqnarray}\label{tam}
\mathcal{J} &\approx& \frac{\hat{z}\varepsilon_{0}}{2\omega}\int \left(\left|E_{L}\right|^{2} -\left|E_{R}\right|^{2}\right)\sqrt{-g}dV\\ 
&+& \frac{\hat{z}\varepsilon_{0}}{2\mathit{i}\omega}\int \left(\overline{E}_{i}\frac{\partial E^{i}}{\partial\phi}\right)\sqrt{-g}dV \nonumber
\end{eqnarray}
Equation \ref{tam} can then be restated using the above spin tetrad formalism and utilizing the transformations easily seen in $\omega = u_{\alpha}k^{\alpha}$ for redefining the angular frequency and the transformation of the unit direction $\hat{z} = e^{\nu}_{3}\gamma_{\nu}$ which is parallel to the spin axes $\hat{z} || \gamma_{+}$. We can further make a substitution for the electric field as a component of the electromagnetic field tensor $F_{\alpha\beta}$
\begin{subequations}
\begin{eqnarray}
E_{i} &\to& E_{\alpha} = F_{\alpha\beta}u^{\beta}\\
E^{*}_{i} &\to& \overline{E}_{\alpha} = \overline{F}_{\alpha\beta}u^{\beta}
\end{eqnarray}
\end{subequations}
with $\left|E_{L}\right|^{2} -\left|E_{R}\right|^{2} \to S_{3}$, giving a redefined function for the total angular momentum of the jet as
\begin{eqnarray}\label{reltam}
\mathcal{J} &\approx& \underbrace{\frac{\varepsilon_{0}}{2u_{\alpha}k^{\alpha}}\int e^{\nu}_{3}\gamma_{\nu}\left(S_{K=3}\right)\sqrt{-g}dV}_{SAM}\\ 
&+& \underbrace{\frac{\varepsilon_{0}}{2\mathit{i}u_{\alpha}k^{\alpha}}\int e^{\nu}_{3}\gamma_{\nu} \left(\overline{F}_{\alpha\beta}u^{\beta}\frac{D \left(F_{\alpha\beta}u^{\beta}\right)}{D\phi}\right)\sqrt{-g}dV}_{OAM}.\nonumber
\end{eqnarray}
Looking at this equation, we can see that a generalization can be made for the SP $(S_{K=3}) = \mathit{i}J_{\alpha\beta}\left(e^{\alpha}_{1}e^{\beta}_{2} - e^{\alpha}_{2}e^{\beta}_{1} \right).$ in the spin part of the total angular momentum. Considering a generality for the case of a partially polarized state we can move to redefine the polarization tensor in $J_{\alpha\beta}$ in terms of the degree of total polarization, $d_{P}$, within the tetrad frame
\begin{eqnarray}
J_{AB} &=& \frac{1}{2}\left(1-d_{P}\right)\delta_{AB} + d_{P}E_{A}\overline{E}_{B}\\
&=& \frac{1}{2}\left(1-d_{P}\right)\delta_{AB} + d_{P}F_{AC}u^{C}\overline{F}_{DB}u^{D}
\end{eqnarray}
and extracting out the degree of circular polarization from this expression allows us to express the SP in the general circularly polarized case, projected on to the tetrad frame,
\begin{equation}
S_{circ} = S_{K=0}\left(d_{C} - \sum_{K}\tilde{S}_{K}\right)
\end{equation}
where $\tilde{S}_{K}$ are the normalized tetrad projected SP with respect to the intensity, $I$. Note that here it is to be observed that $K$ is not a coordinate index on the SP, but merely a label counting the group of parameters $K=(0,1,2,3)$. Making this substitution back into the approximation for the SAM we have
\begin{equation}\label{tsam}
\mathcal{J}_{SAM} \approx \frac{\varepsilon_{0}}{2u_{\alpha}k^{\alpha}}\int e^{\nu}_{3}\gamma_{\nu} S_{circ}\sqrt{-g}dV
\end{equation}
By definition this expression only accounts for the limited degree of freedom of the spin angular momentum generated with respect to the circular polarization. The OAM is then related to an elliptical polarization that may exist if the jet has an extra degree of freedom in its emission where one can observe a departure from the rotational symmetry about the azimuthal axis of the black hole. For this work, we restrict our discussion of the angular momentum associated with the polarization to only considering the contribution from SAM and not the total angular momentum, $\mathcal{J}_{tot}$. The continuation of this work will explore the elliptical polarization due to the transfer of OAM from the central black hole and it's precession around the axis of rotation. 
\\

\subsection{Spin Current}
Looking the relationship between relativistic angular momentum and the geometry of spacetime that is prescribed in this work, one can then construct a \textit{spin tensor} that represents the components of SAM as it relates to the energy-momentum of the object of interest. Here we will focus on the black hole's SAM as it relates to the net spin of a relativistic jet. This can be done by first describing the flux of matter at the boundary of the event horizon area as it relates to the rotation of the black hole. The scalar value of angular momentum of an ideal force-free fluid measured at spatial infinity is
\begin{equation}
J_{matter} = -\int_{\Omega} T^{\alpha}{}_{\beta}R^{\beta}d\Sigma_{\alpha}
\end{equation}
where $R^{\beta} = (0,0,0,\phi)$ is an azimuthal killing vector and $d\Sigma_{\alpha}$ is a surface element on the boundary of the horizon area, $\Omega$, intersecting the horizon across a 2-surface $\partial\omega$ coinciding in a global structure at spatial infinity $\partial\Omega_{\infty}$. We can also write down the scalar value of the angular momentum due to the rotation of the black hole as
\begin{equation}
J_{H} = - \frac{1}{8\pi}\int_{\partial\omega} \nabla_{\beta}R^{\alpha}d\Sigma_{\alpha\beta}.
\end{equation}
We can expand the 2-surface element, $d\Sigma_{\alpha\beta} = \mathit{l}_{[\alpha}\eta_{\beta]}dA_{H}$ and give the representation for the tangent vector, $\mathit{l}_{\alpha}$
\begin{equation}
\mathit{l}_{\alpha} = k_{\alpha} + \omega_{H}R_{\alpha}
\end{equation}
This combines the time-like killing vector, $k_{\alpha}$, as seen in preceding sections, and the azimuthal killing vector scaled by the angular velocity of the horizon. The vector $\mathit{l}_{\alpha}$ behaves like a killing vector tangent to the horizon in the plane of rotation, and $\eta_{\alpha}$ behaves as a unit vector normal to the horizon 2-surface. Such that
\begin{eqnarray}
J_{tot} = &-&\int_{\Omega} T^{\alpha}{}_{\beta}R^{\beta}l_{[\alpha}\eta_{\beta]}dA_{H}\\ 
&-& \frac{1}{8\pi}\int_{\partial\omega} \nabla_{\beta}R^{\alpha}l_{[\alpha}\eta_{\beta]}dA_{H}\nonumber
\end{eqnarray}
describes the total angular momentum of the matter-black hole system. 

If we consider electromagnetic characteristics of this approximation, we can then focus on the coupling of electromagnetic fields to the curvature of spacetime by introducing the components of the electromagnetic energy-momentum tensor governing the density and flux of linear momentum in the propagating fields. For a rotating spacetime we write down an associated angular momentum 3-form as the antisymmetric product of this energy-momentum and a vector orthogonal to the axis of rotation
\begin{eqnarray}
L_{\gamma\mu\nu} &=& \mathit{z}_{[\gamma} T_{\mu]\nu}\\
&=& \mathit{z}_{\gamma}T_{\mu\nu} - \mathit{z}_{\mu}T_{\gamma\nu}.\nonumber
\end{eqnarray}
It is clear to see that upon examination of this equation for the angular momentum density one can incorporate the total current, $J_{tot}$ given above by adding to $L_{\gamma\mu\nu}$ a corresponding term that allows us to describe the angular momentum as a Noether current at a point $(\bf{r})$ about some axis perpendicular to the origin of rotation. This can be written as an expansion of the Lie bracket $[\mathit{z}^{\gamma}, T^{\mu\nu}]$ for some fixed time $\tau=0$, for which
\begin{equation}
L^{\gamma\mu\nu}_{\textbf{r}}(\textbf{z}) = L^{\gamma\mu\nu}_{\textbf{0}}(\textbf{z}) + \left[\mathit{r}^{\gamma}T^{\mu\nu}(\textbf{z}) - \mathit{r}^{\mu}T^{\gamma\nu}(\textbf{z})\right]
\end{equation}
Utilizing this definition of a conserved current rotating about the point $(r)$, a more general representation for the SAM can be given as
\begin{equation}\label{samdensity}
\mathcal{S}^{\gamma\mu\nu}(\textbf{z}) = L^{\gamma\mu\nu}_{\textbf{0}}(\textbf{z}) + \left[\mathit{z}^{\gamma}T^{\mu\nu}(\textbf{z}) - \mathit{z}^{\mu}T^{\gamma\nu}(\textbf{z})\right]
\end{equation}
which represents the SAM about a fixed point $(\textbf{z})$ that is the current at $(\textbf{z})$, rotating about $(\hat{z})$. From here, one can make the statement that the SAM is a conserved quantity for a static extended body in 4 spacetime dimensions implying that there is no loss of angular momentum from the accreting black hole. As we have seen, this is not the case when considering AGN that have energetic relativistic jets formed due to the flux of angular momentum. With this in mind, the \textit{spin current} is a quantity that retains some form of global symmetry with respect to the total angular momentum, but is not considered a \textit{Noether Current} in the sense that the divergence of the spin current is nonzero. This can be seen by the continuity equation
\begin{eqnarray}\label{torque}
\nabla_{\nu}\mathcal{S}^{\gamma\mu\nu} &=& \nabla_{\nu}L^{\gamma\mu\nu}_{\textbf{0}}(\textbf{z})\\ 
& +& \nabla_{\nu}\left[\mathit{z}^{\gamma}T^{\mu\nu}(\textbf{z}) - \mathit{z}^{\mu}T^{\gamma\nu}(\textbf{z})\right]\nonumber
\end{eqnarray}
where $\nabla_{\nu}L^{\gamma\mu\nu}_{\textbf{0}}(\textbf{z}) = 0$ and implies that $T^{\gamma\mu}$ is no longer symmetric in its components. 
\begin{equation}
\nabla_{\nu}\mathcal{S}^{\gamma\mu\nu} = T^{\mu\gamma} - T^{\gamma\mu}
\end{equation}
This also implies that the divergence of the spin current gives an associated nonzero torque density that quantifies the transformation of OAM to SAM over many cycles.

A statement can be made for this spin current to be related back to the angular momentum of the black hole with the 3-surface of the horizon as a boundary $\partial\Omega$ on the extent of the spin, restricting ourselves to the physics contained within the ergosphere and up to the boundary of the horizon. Integrating the spin current with respect to this boundary condition provides the aforementioned divergence as
\begin{equation}\label{torque}
\int_{\Omega} \nabla_{\nu}\mathcal{S}^{\gamma\mu\nu}(\textbf{z})\sqrt{-g}d^{n}\mathit{z}_{H} = \oint_{\partial\Omega} \mathcal{S}^{\gamma\mu\nu}(\textbf{z}) \eta_{\nu}d^{n-1}\mathit{z}_{H}.
\end{equation}
According to equation \ref{torque}, and exploiting the divergence theorem, the  nonzero density is validated by the right hand side of this equation. It can be further stated that the torque density is then a smooth function of $\textbf{z}$ at $\tau = 0$ in the $\hat{z}$ direction. Consider the spin current averaged over many cycles (or periods, T) for the case of a relativistically accreting black hole, over spacelike hypersurfaces the average spin is then
\begin{equation}
\left<\mathcal{S}^{\gamma\mu\nu}(\textbf{z}) \right> = 2\omega\int_{\tau=0}^{\tau=T} L^{\gamma\mu\nu}_{\tau_{N}}(\textbf{z})d\tau
\end{equation}
Integrating the angular momentum density over the proper time allows one to relate the average energy-momentum contained within the jet $\left<T^{\mu\nu}(\textbf{z})\right>$ back to the rotation of the black hole using $J_{matter}$ such that, again, the divergence represents the transformation of OAM to SAM
\begin{equation}
\nabla_{\beta}\mathcal{S}_{\mu\nu} = \oint_{\partial\omega} \mathcal{S}_{\mu\nu\alpha}(\textbf{z}) l_{[\alpha}\eta_{\beta]}dA_{H}.
\end{equation}
\begin{figure}[h]\label{killing}
\centering
\includegraphics[scale=0.25]{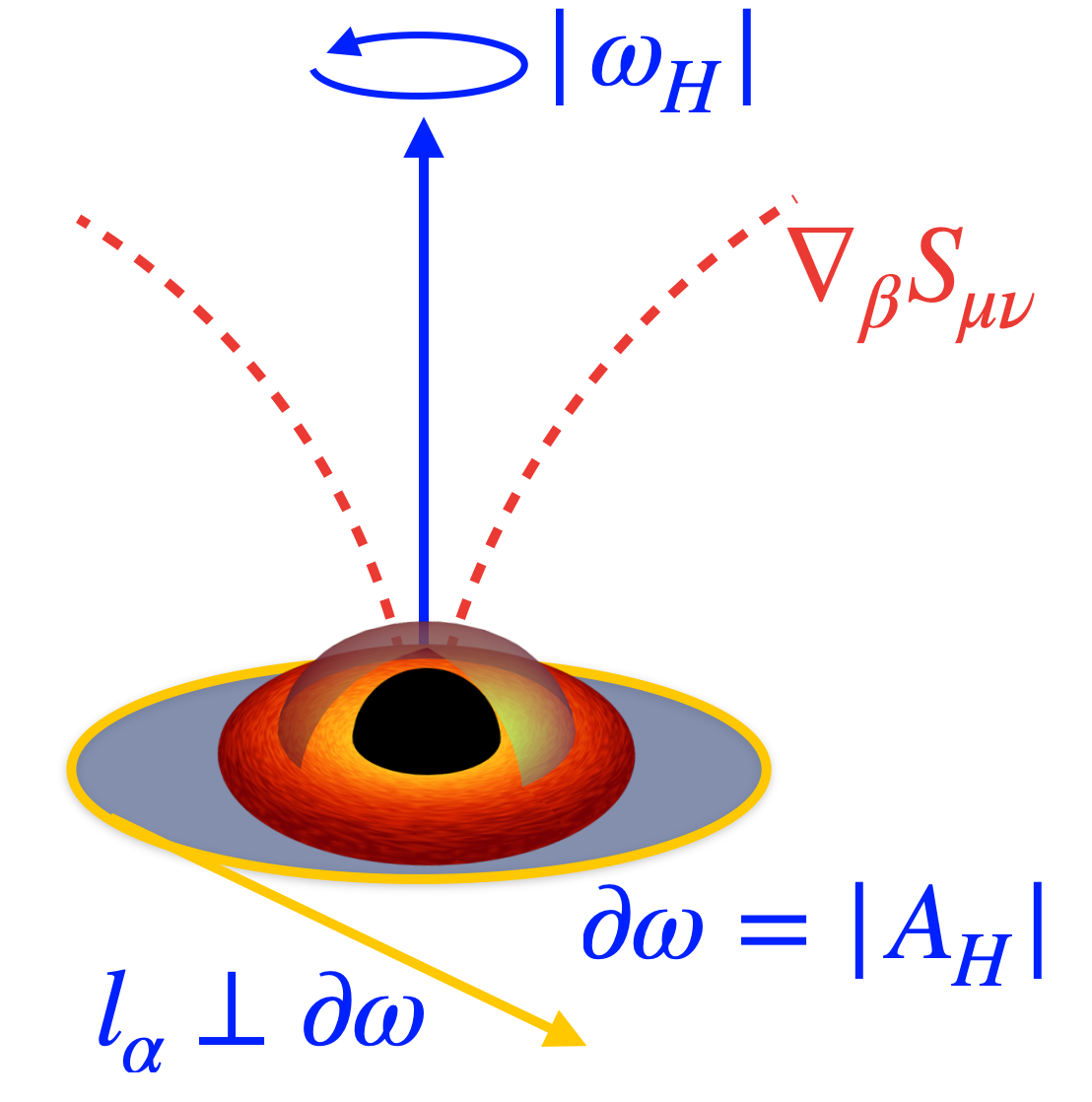}
\caption{Visualization of the divergence of the spin current on the horizon boundary and the rotational killing vector, $\mathit{l}_{\alpha}$.} 
\end{figure}
\\
Now, let $\mathcal{J}_{SAM}$ be the SAM of the jet in the $e_{3}^{\nu}\gamma_{\nu}$ direction and let $\Re\{\mathcal{S}^{\gamma\mu\nu}\} \in \mathbb{R}^{4}$. It can be seen that $\mathcal{J}_{SAM}$ is a component of $\left<S^{\gamma\mu}\right>$ in the frame parallel to $e_{3}^{\nu}\gamma_{\nu}$.
\begin{equation}\label{spinpolar}
\left<S^{03}\right> = \mathcal{J}_{SAM} \approx \frac{\varepsilon_{0}}{2u_{\alpha}k^{\alpha}}\int e^{\nu}_{3}\gamma_{\nu} S_{circ}\sqrt{-g}dV
\end{equation}

We can conclude this by making the simplified transformation to the spin tetrad frame resulting in a complete description for the \textit{Spin states per unit volume}
\begin{equation}
\mathcal{S}_{LKM} = \mathcal{S}_{\delta\mu\nu}\gamma_{L}^{\delta}\gamma_{K}^{\mu}\gamma_{M}^{\nu}.
\end{equation}
\\ 
\section{Discussion}
\subsection{The Observation-Theory Relationship}
As we have seen a relativistic jet described as a beam of light carries linear momentum, and thus is influenced by an appreciable amount of external angular momentum in both the non-relativistic and relativistic regimes. This angular momentum would then be dependent on the origin of an associated coordinate system, owing to the intrinsic gauge dependence of angular momentum in fundamental physics descriptions. If we then proceed to describe BL Lac and FSRQ blazars as energetic point-sources, we can infer the physical characteristics of the jet emission as relativistic beams transported across galactic distances. These point sources should then inherently carry a rotational symmetry corresponding to rotated field lines with respect to the host black hole. Thus, removing some of the mystery of the physical mechanisms that causes some jets to twist and carry a proportionate amount of angular momentum from the black hole. It is then intuitive to think about how one can infer the mechanisms causing such polarizations in the observed spectra. Observations of blazars and radio-loud sources have shown that polarization states exist in the spectra from these sources \cite{cp_blazar2021}. Recent observations of x-ray and gamma-ray sources have eluded to the need for more missions that carry polarimetry capabilities \cite{crabpolarimetry, m87spin, Rani2019,cp_blazar2021,Rieger2005}. 

A review of observed features of relativistic jets from radio-loud AGN provides contact with this present work in regards to stitching the mathematical general relativistic theory to observable variables for describing the physics governing the launch mechanisms for these jets. This work contributes to the fundamental description of jets emanating from strong sources of gravity (SMBH) and the influence those sources have on the surrounding environment that the jet is emitted in. Relativistic jets, in and of themselves, are extended objects such that they act as beacons of light being pushed out from the edges of the outer event horizons. These objects can extend over large distances ($kpc\lesssim r \lesssim Mpc$ scales) and interact with other matter-energy in the surrounding environment near the black hole \cite{darkagn1,Harris2006,Blandford2019}.

The results of section \ref{sec: amflux} elucidates that there is a relationship between the continuum nature of relativistic jets and the observed polarization fractions that are observed with respect to the angular momentum coupling of matter to the central black hole. Within the derivations outlined above, equation \ref{reltam} provides a description of how the polarization states are influenced by the angular momentum coupling within the vicinity of the event horizon as measured from within a detector or laboratory frame. We can see that in equation \ref{tsam} the SAM becomes a function of the black hole spin parameter $\alpha$ and the polar coordinates $(r,\theta)$, further stitching together such relationships between the black hole spin and the emitted angular momentum flux contained in the jet.  Equation \ref{spinpolar} can be expanded to include the parameters form the Kerr metric. This can be seen by expanding the azimuthal frame tetrad 1-form $e^{\nu}_{3}\gamma_{\nu}$ in the spin current tensor below,
\begin{widetext}
\begin{eqnarray}
\left<S^{03}\right> &=&  \frac{\varepsilon_{0}}{2u_{\alpha}k^{\alpha}}\int e^{\nu}_{3}\gamma_{\nu} S_{circ}\sqrt{-g}dV\\
&=& \frac{\varepsilon_{0}}{2 \omega_d}\int\left[-\left(r^2+\alpha ^2+\frac{M_{BH} r\alpha^2}{(r^{2}+\alpha^{2}\cos^{2}\theta)^{2}}\sin^2\theta \right)\sin^2\theta\right]S_{circ}\sqrt{-g}dV \nonumber
\end{eqnarray}
\end{widetext}
We can then make the restraint that relevant distances when considering this process is within the region bounded by the rotation of the event horizon at $r > z_{H} = \frac{1}{2}\left[r_{s} - \sqrt{r^{2}_{s} -4\alpha^{2}}\right]$. The circular polarization parameter $S_{circ}$, and subsequent degree of circular polarization in the tetrad frame $d_{C}$, can then be expanded as well in terms of the black hole parameters using the metric function $\rho(\alpha, r) \text{and } M_{bh}$  as
 \begin{widetext}
 \begin{eqnarray}
 S_{circ} &=&  J_{\alpha\beta}\left(e^{\alpha}_{1}e^{\beta}_{1} + e^{\alpha}_{2}e^{\beta}_{2} \right)\left(d_{C} - \sum_{K}\tilde{S}_{K}\right)\\
 &=& \frac{1}{2}\left<c^{2}F_{10}\overline{F}_{10}\right>\left(\frac{r^{2}+\alpha^{2}\cos^{2}\theta}{r^{2}-M_{bh}r +\alpha^{2}}\right) +\left<c^{2}F_{20}\overline{F}_{20}\right> \left(-\rho^{2}\right)\left(d_{C} - \sum_{K}\tilde{S}_{K}\right) \nonumber
 \end{eqnarray}
\end{widetext}
\begin{widetext}
\begin{equation}
d_{C} = \frac{\left[-\frac{1}{4}\left[\left(\left<c^{2}F_{10}\overline{F}_{20}\right>\left(-\sqrt{\frac{\rho^{2}}{\Delta}}\sqrt{-\rho^{2}}\right)\right)^{2} - \left(\left<c^{2}F_{20}\overline{F}_{10}\right>\left(-\sqrt{\frac{\rho^{2}}{\Delta}}\sqrt{-\rho^{2}}\right)\right)^{2}\right]\right]^{1/2}}{\frac{1}{2}\left<c^{2}F_{10}\overline{F}_{10}\right>\left(\frac{r^{2}+\alpha^{2}\cos^{2}\theta}{r^{2}-M_{bh}r +\alpha^{2}}\right) +\left<c^{2}F_{20}\overline{F}_{20}\right> \left(-\rho^{2}\right)} 
\end{equation}
\end{widetext}
It is the intricate relationship between black hole spin, electromagnetic stokes parameters, and the distribution of matter at high energies that is the essence of this work. This paper supports the fundamental understanding of how the spin angular momentum of black holes can have a direct influence on the emission of charged particles at sufficient energy levels. As part of a larger body of work, which is on-going, this paper provides derives a portion of the mathematical theory that is needed to lay the foundation for providing predictions or constraints on observational parameters of jets. Lastly, it is no small task to bridge the gap between the physics of black holes, per the mathematical theory of general relativity, to the archived observational data presented in the reviews of AGN catalogues. This work contributes to the overall notion that black holes (curved spacetime) influence the motion and emission of particles with sufficient energy.
\\
\section{Conclusion}
As we've seen in this work, there are more clues and properties to uncover pertaining to the dynamics of relativistic jets, their polarizations and formation. A full theoretical description of the SAM-OAM coupling for polarization states, where the rotational gauge freedom is all but used, will prove to be a more robust theory. Of note, the resulting derivation of the above theory gives a constrained view of the relationship between the distribution of matter and the rotation of the central black hole. Thus contributing to the supporting scientific predictions for future x-ray and gamma-ray polarimetry missions surveying AGN. A broader discussion on the flux of total angular momentum from the central black hole requires extending our understanding of energy-momentum emission characteristics from Kerr black holes. Focusing on the theoretical aspects of jet formation mechanisms, there are still fundamental questions that continue to remain unresolved. It is evident that further evaluation of the remaining SP that were not explored in this work (linear and elliptical states) is needed to have a complete description of their respective symmetries.

\section{Acknowledgments}
This material is based upon work supported by NASA Goddard Space Flight Center under award number 80GSFC21M0002. I'd like to thank Dr. Rita Sambruna and Dr. Demos Kazanas for their valuable discussions and insights on this topic. In addition, I'd like to thank the reviewers for their informative recommendations.


\bibliographystyle{apsrev4-2}
\typeout{}
\bibliography{REF_full.bib}

\end{document}